\pdfoutput=1

\documentclass[screen,nonacm]{acmart}
\AtBeginDocument{%
  \providecommand\BibTeX{{%
    \normalfont B\kern-0.5em{\scshape i\kern-0.25em b}\kern-0.8em\TeX}}}

\usepackage[suppress,showdeletions]{color-edits}
\usepackage{cleveref}
\usepackage{multirow}
\usepackage{enumitem}
\usepackage{array}
\usepackage{svg}

\newcommand{\Crefp}[1]{(\Cref{#1})}
\addauthor{samson}{teal}
\addauthor{araz}{red}
\addauthor{kathy}{purple}

\setcopyright{acmcopyright}
\copyrightyear{2022}
\acmYear{2022}
\acmDOI{-}
\newcommand{\etal}{et al.}

\begin{document}

\title{The Risks of Machine Learning Systems}


\author{Samson Tan}
\email{samson.tmr@comp.nus.edu.sg}
\orcid{}
\affiliation{
  \institution{Salesforce Research Asia, and School of Computing, National University of Singapore} 
 }

\author{Araz Taeihagh}
\email{spparaz@nus.edu.sg}
\orcid{https://orcid.org/0000-0002-4812-4745}
\affiliation{
\institution{Lee Kuan Yew School of Public Policy and CTIC, National University of Singapore}
}

\author{Kathy Baxter}
\email{kbaxter@salesforce.com}
\orcid{}
\affiliation{
  \institution{Office of Ethical \& Humane Use, Salesforce}
}


\begin{abstract}
The speed and scale at which machine learning (ML) systems are deployed are accelerating even as an increasing number of studies highlight their potential for negative impact. There is a clear need for companies and regulators to manage the risk from proposed ML systems before they harm people. To achieve this, private and public sector actors first need to identify the risks posed by a proposed ML system. A system’s overall risk is influenced by its direct and indirect effects. However, existing frameworks for ML risk/impact assessment often address an abstract notion of risk or do not concretize this dependence. Keeping discussions of risk at an abstract level puts the onus of defining them on the assessor, which may result in incomplete and inconsistent assessments.

We propose to address this gap with a context-sensitive framework for identifying ML system risks comprising two components: a taxonomy of the first- and second-order risks posed by ML systems, and their contributing factors. First-order risks stem from aspects of the ML system, while second-order risks stem from the consequences of first-order risks. These consequences are system failures that result from design and development choices. We explore how different risks may manifest in various types of ML systems, the factors that affect each risk, and how first-order risks may lead to second-order effects when the system interacts with the real world. 

Throughout the paper, we show how real events and prior research fit into our Machine Learning System Risk framework (MLSR). MLSR operates on ML systems rather than technologies or domains, recognizing that a system's design, implementation, and use case all contribute to its risk. In doing so, it unifies the risks that are commonly discussed in the ethical AI community (e.g., ethical/human rights risks) with system-level risks (e.g., application, design, control risks), paving the way for holistic risk assessments of ML systems.
\end{abstract}

\maketitle


\section{Introduction}
While machine learning (ML) has the potential to improve the quality of our lives, it also has the potential to bring about new harms and exacerbate existing ones. \samsonedit{Much of the ML literature has naturally focused more on the benefits of ML than the harms to society. Therefore, we provide a counter-perspective by categorizing the risks posed by ML systems. For the purposes of discussion, we consider an ML system to be the ML-based component(s) of a software system. In some scenarios, this software system may also interface with a hardware system, such as in an autonomous robot. Since the risks of traditional computer systems have been extensively discussed \cite{sherer1995three,higuera1996software}, we focus our efforts on the new, or increased, risks brought about by their ML components.} Due to existing power structures, ML systems often benefit some demographics at the expense of others \cite{oneil2016weapons,eubanks2018automating,lim2019algorithmic,benjamin2019race,Taeihagh2021}. Recent work and news have highlighted the negative impacts of ML systems on society and the environment \cite{buolamwini2018gender,dastinamazonhiring2018,lim2018autonomous,feathersflawed2019,strubell2019energy,bender2021dangers,lim2019algorithmic,TaeihaghLim2021}. For example, Rousseau \etal\ found GPT-3 \cite{gpt3} to be highly unreliable and downright dangerous for healthcare applications \cite{drgpt}.

This has led to calls and proposals for legislation to regulate the sale and use of ML systems, such as the draft EU AI Act \cite{euaiact2021}, the US’ Algorithmic Accountability Act (AAA) \cite{aaa2019}, and California’s Automated Decision Systems Accountability Act (ADSAA) \cite{adsaa2021}. The EU AI Act focuses on predefined classes of applications deemed either "prohibited" or ``high risk'' (Title III), leaving others to be governed by existing product safety regulations with an exception being transparency obligations for specific classes of applications (Title IV). Presently, it does not provide the guiding criteria for the creation of these lists. In contrast, the AAA and ADSAA outline general criteria for an ML system to be considered ``high risk'', such as whether the system uses personal or demographic-related information. However, none of the three regulations provide guidance on how ``medium" and "lower risk'' applications should be assessed and managed. China's white paper on trustworthy AI goes into more detail on managing teams and the development process, but does not discuss risk-based differentiated measures \cite{caict2022trustworthy}. We argue that accurately characterizing the risks posed by ML systems is crucial to enacting meaningful regulations while not stifling innovation \cite{canes2001leadership,taeihagh2019governing}. We note that the latter does not mean giving businesses a free pass to not find safer, more inclusive solutions and prioritize their profits above all. Instead, organizations must recognize and address the risks of technological disruption. A comprehensive risk taxonomy will enable this, while also helping organizations define internal policies on ML applications not covered by the law.

Although the European Commission and German Data Ethics Commission have published frameworks for risk-based regulation \cite{gdc2019,euaiact2021}, algorithmic impact assessments (IAs) are often used in the literature instead of risk assessments (RAs) \cite{ethicstoolkit2018,catoolkit2018, reisman2018algorithmic,mantelero2018ai,charternz2020,harmsmodel2020}.  IAs primarily focus on the ML system's impacts (e.g., on human rights and the environment) while RAs, in addition, are able to surface direct risks of the system that contribute to its impacts.

Even as IAs and RAs are increasingly mandated by law \cite{euaiact2021,aaa2019,adsaa2021,cadirective2021}, the notion of risk, impact, or harm is ill-defined in most existing IA/RA frameworks and proposed regulations \cite{catoolkit2018,mantelero2018ai,aaa2019,firlej2020regulating,euaiact2021,adsaa2021}. This is likely due to their tendency to focus on algorithmic instead of ML systems (a subset of the former). Consequently, they lack detailed discussions of the risks and risk factors specific to ML systems. This creates loopholes that can be exploited by malicious actors \cite{veale2021demystifying} and risk identifying only easily measurable harms \cite{metcalf21}. Other studies examine in detail the risks of specific applications (e.g., autonomous vehicles) \cite{milakis2017policy,lim2019algorithmic} but many of the highlighted risks are domain-specific and not broadly applicable. We argue that a structured understanding of ML-specific risks is crucial to the development of comprehensive RAs.

Software risk assessment is a well-established practice with taxonomies that overlap with the concerns of ML risk assessment, such as reliability, safety, and security \cite{sherer1995three,higuera1996software}. However, work in this area has been primarily concerned with the system's negative consequences for the \emph{organization developing it}, as opposed to the \emph{affected communities}. While the latter has always existed, the recent proliferation of ML systems has increased both their likelihood and severity, in addition to introducing new risks resulting from learned behavior.

In summary, existing work on algorithmic IA and software RA lacks detailed discussions of ML-specific risks. On the regulation side, upcoming legislation does not provide sufficient guidance on non--high-risk applications and often refer to a vague idea of risk. To advance the field on this front, we develop a Machine Learning System Risk framework (MLSR) of the risks posed by ML systems and their risk factors. MLSR is inspired by  existing work on software risks and tech-related harms, application-specific studies, and the Universal Declaration of Human Rights \cite{higuera1996software,udhr,ethicstoolkit2018,taeihagh2019governing,harmsmodel2020}. MLSR connects the direct risks of a system (first-order risks) to the risks that arise from its interaction with the real world (second-order risks), discussing in detail the factors that contribute to each type of risk. This will help organizations perform holistic risk assessments, devise appropriate mitigation measures, and make project approval/denial decisions, especially when a proposed ML system does not appear immediately dangerous. Finally, an understanding of the risks surrounding ML systems will aid in the creation of appropriate and enforceable standards and regulations. \samsonedit{In this paper, we focus on organizing the risks posed by ML systems into a framework of first and second order harms and leave the \emph{assessment} of these risks to future work.}

\section{Risk and Impact Assessment of Machine Learning Software Systems}
Impact assessments aim identify and monitor the positive or negative consequences of existing and planned projects \cite{vanclay2003international}, while risk assessments focus on the potential \emph{negative} consequences of \emph{planned} projects \cite{rausand2013risk}. Such assessments help us assert some control over the future instead of being subjected to its whims \cite{bernstein1998against}. We use the ISO definition of risk: the consequence of an event combined with its likelihood of occurrence \cite{isorisk}. 

Risk and impact assessments have been (and continue to be) extensively used across industries to evaluate both the risk/impact \emph{of} projects on the environment and affected communities \cite{meetham1945atmospheric,munn1979environmental,fullwood1988probabilistic,willcocks1994risk,efroymson2000ecological,brown2007identifying,esteves2012social,cook2013spent,de2016priam} and the risk \emph{to} a project or system \cite{sherer1995three,soares2001risk,haimes2002risk,peltier2005information,zavadskas2010risk,lavasani2011fuzzy}. Algorithmic IAs evolved from those for social and environmental impact \cite{munn1979environmental,vanclay2003international,mantelero2018ai}, while RAs have a long history of usage for safety- or mission-critical applications, such as nuclear power plants \cite{fullwood1988probabilistic}, hazardous material transportation \cite{verter2001gis}, civil aviation \cite{janic2000assessment}, and civil engineering \cite{zavadskas2010risk}.

In information systems and software development, risks are often studied from the organization's perspective \cite{sherer2004information}; negative consequences are note-worthy if they threaten the organization. For example, Susan Sherer breaks software risk down into three categories of consequences for the organization: development, use, and maintainability \cite{sherer1995three}. Of the twelve risks that make up this taxonomy, only one captures the impact of the developed software on its users, society, or the environment --- safety. Higuera and Haimes present a comprehensive three-level taxonomy of sixty-four software risks \cite{higuera1996software}. Similarly, almost all of these risks are concerned with threats to the project's success, with only two --- safety and human factors --- addressing risks to its users. While the negative impact of software systems on society and the environment is not a new phenomenon \cite{rutkowski2013generational}, the recent proliferation of biased and unsafe ML systems has cast the spotlight on software systems and the algorithms powering them \cite{oneil2016weapons,benjamin2019race}.

Consequently, researchers in the ethical AI and public policy communities have called for the use of IAs \cite{selbst2018disparate,raji2020closing,moss2021assembling} to surface the direct, indirect, and insidious harms that could be caused by an ML system, such as entrenching historical discrimination. Such assessments can be a way of improving accountability in the use, procurement, and development of ML systems by ensuring actors perform due diligence \cite{reisman2018algorithmic,aaa2019,lim2019algorithmic,raji2020closing,euaiact2021}. However, regulatory support is crucial to accountability---it cannot exist without the threat of consequences for non-compliance \cite{bovens2007analysing,taeihagh2021assessing,moss2021assembling}. Reisman \etal\ highlight the difficulties facing governments in assessing the impact of automated decision systems\footnote{Often used as a catch-all synonym for algorithmic, AI, or ML system with varying definitions. We only use this term to accurately represent the references that use it.} (ADSs), in large part due to their ``black-box'' nature and trade secrecy claims \cite{reisman2018algorithmic}. While they provide some guidelines for implementing IAs in the public agency context, they do not identify specific relevant risks or impacts.

In contrast, the Canadian government's ADS IA tool is a questionnaire for assessing socioeconomic and environmental impact, impact on government operations, system complexity, data management, and procedural fairness \cite{catoolkit2018,caaiamedium}. While  comprehensive, it lacks a well-defined taxonomy of impacts to guide users and does not examine the system in detail \cite{lemayaiamedium2019,metcalf21}. On the other hand, Microsoft's Harms Modeling framework comprises a comprehensive taxonomy of thirty-eight tech-related impacts and example questions to assess each one \cite{harmsmodel2020}. Krafft \etal\ presented the Algorithmic Equity Toolkit with an IA worksheet for community members to assess the impact of a government-deployed ADS on them \cite{krafft2021action}. New Zealand's Algorithm Charter appears to bridge the gap between IAs and RAs by defining impact as a component of risk, 
but does not define the terms ``algorithm'', ``risks'' and ``impacts''. The Ethics and Algorithms Toolkit provides a step-by-step checklist that operationalizes algorithmic risk as impact on people and property, data use, level of accountability, and bias \cite{ethicstoolkit2018}. Impact assessments may also be performed post-hoc and informally by external/independent auditors to highlight the flaws in already deployed ML systems, such as in the Gender Shades study \cite{buolamwini2018gender}.
However, it is important to acknowledge that post-hoc analysis is often challenging, especially for ML platforms that do not have access to customer data to detect if their customers are using the system to cause harm. Furthermore, it may be difficult to observe some vulnerable communities to determine if they are disparately impacted unless user demographic data is collected. Despite their usefulness, existing IAs risk only operationalizing ``impact'' in ways amenable to computational evaluation. This will lead to a loss of trust when they inevitably fail to capture pertinent harms \cite{metcalf21}.

To summarize, research on software RAs largely focus on the risk to the organization, as opposed to the risk to society or the environment. Existing work on algorithmic IAs identifies the negative consequences of algorithmic systems to society and the environment but focuses primarily on discrimination risk or does not link them to specific system development choices. \samsondelete{We focus on risk assessment in this paper since an ML system's impacts can be framed as risks but its direct risks cannot be framed as impacts.} An organization will require such traces to address the concerns identified by IAs. Presently, proposed legislation also tends to address ``risks'' and ``impacts'' at a high level, which need to be operationalized in greater detail to be useful. In contrast, MLSR documents the various failure modes of an ML system and their contributing factors, and connects them to its impacts on society and the environment. 

\begin{figure}
    \centering
    \includesvg[width=\textwidth]{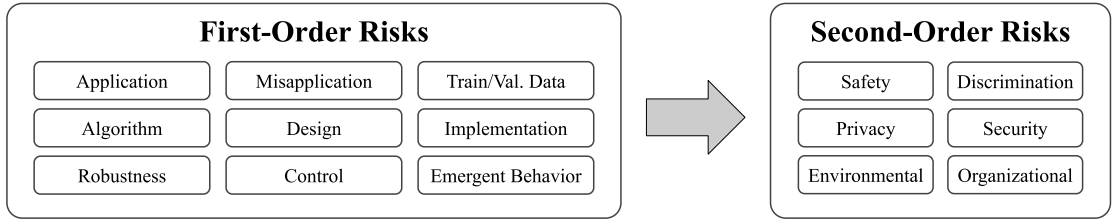}
    \caption{Overview of the ML System Risk (MLSR) framework. First-order risks stem directly from the machine learning system \Crefp{sec:first_order} and their consequences lead to second-order risks when the system interacts with the real world \Crefp{sec:second_order}. We summarize them in \Cref{tab:first_order,tab:second_order} \Crefp{app:tables}.}\label{fig:main}
    \vspace{-1em}
\end{figure}

\section{A Taxonomy of Machine Learning System Risks}
The risks posed by ML systems can be categorized into first-order, second-order, and beyond. We explore first- and second-order risks in the following sections but acknowledge the presence of higher-order risks due to knock-on effects. While some of the risks described below may apply to all software systems, we focus on the aspects specific to ML systems. \samsonedit{Hence, we will refer to our framework as the Machine Learning System Risk framework (MLSR).}

To construct this taxonomy, we used algorithmic impacts as a starting point, with safety, discrimination, and environmental risks \cite{ethicstoolkit2018,strubell2019energy,harmsmodel2020}. We then included software risks pertinent to ML systems, such as design, implementation, safety, privacy, security, and organizational risks \cite{sherer1995three,haimes2002risk}. 
Separately, we surveyed the ML literature and identified common themes such as training/validation data, algorithm, robustness, design, implementation, privacy, security, and emergent behavior risks. We supplemented the above with reporting on ML-related incidents and professional experience, identifying more risks such as application, misapplication, and control risks. Concurrently, we also identified factors that affect each risk. Finally, we grouped them into first- and second-order risks using the following criteria: First-order risks directly arise from the choices made during the ML system's conception, design, and implementation, and relate to the ways it can fail. The consequences of these choices lead to \emph{second-order risks} when the ML system interacts with the world. Second-order risks, hence, relate to the \emph{impact} of first-order consequences on the real world. \samsonedit{Although differentiating the risks into two orders increases MLSR's complexity, it allows practitioners to safely exclude a second-order risk from their list of concerns if the ML system is not vulnerable to the associated first-order risk.}

\section{First-Order Risks}\label{sec:first_order}
First-order risks can be generally broken down into risks arising from intended and unintended use, system design and implementation choices, and properties of the chosen dataset and learning components. \emph{Note: This is an unordered list.}

\subsection{Application risk}\label{app_risk} 
This is the risk posed by the intended application or use case. It is intuitive that some use cases will be inherently ``riskier'' than others (e.g., an autonomous weapons system vs.\ a customer service chatbot).

\paragraph{Application domain}
As alluded to above, the intended purpose of the ML system can be a major risk factor, holding all other variables constant. Other than the specific use case, the domain could also contribute to the application risk. For example, it is intuitive that the negative consequences are more severe for an image classification system used to aid melanoma diagnoses than one used for Lego brick identification.

\paragraph{Consequentiality of system actions}
The impact of the ML system's actions on the affected community members is another important factor in the system's application. For example, a slightly inaccurate automated text scoring system carries relatively minor consequences if used only for providing feedback on ungraded homework, compared to being used for grading school assignments. While inaccuracies in the latter use case may affect a student's annual ranking, it carries a lower risk compared to using the same system to grade national exams that determine a student's future, where even minor inaccuracies can unfairly impact their ability to enter their desired university or major \cite{feathersflawed2019}.

\paragraph {Protected populations impacted} Most societies have special protections for certain population groups such as children, the elderly, disabled, or ethnic minorities. For example, in the US, the Child Online Privacy Protection Act imposes stricter requirements on operators of websites or online services directed to children under 13 years of age \cite{coppa1998}. Similarly, some social groups may be more vulnerable to the negative impacts of an ML system and lower thresholds for harm may therefore be necessary for them. The US Federal Trade Commission has warned of penalties against companies that sell or use biased AI systems that harm protected groups \cite{ftcwarning2021}. 

\paragraph{Effect on existing power differentials and inequalities}
Use cases that entrench or amplify power differentials between the organization employing the system and the affected population should be assigned a higher risk from a human rights perspective. This can take the form of increased surveillance
, which increases the organization's power over the public but not vice-versa. Other applications may amplify systemic inequalities due to the ease, scale, and speed with which predictions can now be made \cite{eubanks2018automating}. Additionally, the act of codifying it in a potentially black-boxed ML system may entrench these learned biases when humans fail to question their predictions \cite{wordofmachine2020}.

\paragraph{Scope of deployment environment}
A system operating in an open environment, such as the outdoors, will often have to account for more uncertainties than in a closed one, such as an apartment. Consequently, there is a higher likelihood of failure in the former. For example, an autonomous cleaning robot deployed in a park will be exposed to a significantly more diverse range of inputs than one used in an apartment. In the latter, the system does not need to handle significant changes in weather conditions and seasons. Additionally, the ability to navigate uneven and unstable terrain will likely be less critical for an indoor cleaning robot compared to one deployed in a park. We refer to this ``openness'' as the deployment environment's \emph{scope}: a wider scope presents more potential points of failure and, therefore, a higher risk.

\paragraph{Scale of deployment}
The scale of a use case will also significantly affect its risk. For example, a system that affects a community of 42 will likely have a lower upper bound of negative consequences compared to being deployed worldwide.

\paragraph{Presence of relevant evaluation techniques/metrics}
Although held-out accuracy is commonly used to evaluate ML models developed for research, this assumes that the training distribution and the deployment environment's distribution are identical. Such evaluation will be insufficient for ML systems meant to be used in the real world since this assumption is often violated. The result is poor system robustness to distributional variation with various second-order consequences (see \Cref{robustness_risk}). Therefore, any evaluation of an application's risk must consider the availability of metrics to evaluate performance on the dimensions relevant to the application or deployment environment \cite{tan-etal-2021-reliability}. For example, a task-oriented chatbot should not only be evaluated using the success rate of the held-out validation set, but also its ability to cope with misspellings, grammatical variation, and different dialects, and generate sentences in the appropriate register. The lack of appropriate metrics reduces the ability to detect such flaws before deployment and increases the risk of negative consequences. Similarly, it is difficult to predict the impact of a risk on the real world. For example, group-level F1 scores for a face recognition system are not indicative of the magnitude of the system's impact on an individual when it is wrong in the real world (e.g., the consequences of arresting a wrongly identified but innocent minority \cite{cvarrest2020}).

\paragraph{Optionality of interaction}
The ability to opt-out of interacting with or being affected by an ML system can limit its negative impacts on a person. For example, choosing to interact with a human customer service agent rather than a chatbot \emph{may} reduce the risk of being misunderstood if the chatbot has not been specifically trained on the customer's language variety. Inversely, being unable to opt-out of the interaction may increase the likelihood and frequency that an individual will experience negative consequences resulting from the ML system. For example, replacing human agents with automated ones as interfaces to essential services may unintentionally prevent the underprivileged from using them due to linguistic barriers. This is a real possibility when the agents have trained on the prestige variety of a language, but the people most in need of access to social welfare services only speak a colloquial variety. 

\paragraph{Accountability mechanisms}
From an organizational perspective, mechanisms that hold the actors accountable for the systems they build reduce the likelihood of negative consequences. For example, an organization might create explicit acceptability criteria, such as comparable accuracy across social groups, reward engineers for meeting these criteria, and block deployment when the system falls short. However, this will only work when acceptance criteria are not in conflict (e.g., engineers being rewarded more for increased user engagement than meeting an acceptable bias threshold). 

\paragraph{Stakeholders' machine learning literacy}
To give useful feedback and seek remediation, the affected community member might require basic knowledge of how ML systems work and the ways they could be impacted. For example, someone unaware of how recommendation algorithms work (or even the existence of such algorithms) may be unable to appreciate the extent to which their political views are influenced by their consumption of social media and video streaming sites \cite{bessi2016users,gil2017effects,beam2018facebook,rochert2020homogeneity}.\footnote{``Any sufficiently advanced technology is indistinguishable from magic'' --- Arthur C. Clarke \cite{clarke2013profiles}.} The affected individual will hence be unaware that they are in an echo chamber, resulting in an inability to break free or give appropriate feedback to the product developers \cite{jeon2021chamberbreaker}. Research has also shown a person's knowledge of AI to affect their interpretation of machine-generated explanations \cite{ehsan2021explainable}.

\subsection{Misapplication risk}
This is the risk posed by an ideal system if used for a purpose/in a manner unintended by its creators. In many situations, negative consequences arise when the system is not used in the way or for the purpose it was intended, and can be thought of as being ``misapplied''. An example is a semi-autonomous vehicle being used as if it were fully autonomous, with the driver taking their hands off the wheel or even leaving the driver's seat while the vehicle is in motion \cite{teslabackseat2021}.

\paragraph{Ability to prevent misuse.}
The ability to prevent misuse before it occurs significantly reduces misapplication risk. In the case of autonomous vehicles, the car might be programmed to automatically slow to a stop if individuals remove their hands from the wheel or if there is a significant weight decrease in the driver's seat while the car is in motion. However, while such failsafes significantly reduce risk, they do not entirely eliminate it since they can be bypassed \cite{teslaautopilotdriver2021}.

\paragraph{Ability to detect misuse}
Being able to detect if the ML system is being used for unintended purposes is crucial to preventing misuse. This can take the form of a component that alerts the organization when a user tries to process inputs with features that match those belonging to prohibited applications (e.g., using a computer vision system for physiognomic purposes), or detect prohibited actions (e.g., leaving the driver's seat when the semi-autonomous vehicle is in motion). Merely relying on whistleblowers and journalists to detect misuse will likely result in the vast majority of misuses going undetected. The detection method's efficacy would, therefore, inversely affect the misapplication risk.

\paragraph{Ability to stop misuse}
Assuming it is possible to detect misapplication, the next factor in managing this risk is an organization's ability to stop misuse once it has been detected. For example, the ability to detect if a customer is using a computer vision system for an unacceptable application (e.g., face recognition for predictive law enforcement) and terminate their access will significantly lower the likelihood of the system being used for such purposes. This is directly related to the system's \emph{control risk} (see \Cref{control_risk}). Being able to instantly shut the system down or terminate the user's access will lower the likelihood and severity of negative consequences stemming from misuse, compared to a delayed or non-response, and could be the difference between life and death for the people affected by the system.

\subsection{Algorithm risk}\label{algo_risk}
This is the risk of the ML algorithm, model architecture, optimization technique, or other aspects of the training process being unsuitable for the intended application. Since these are key decisions that influence the final ML system, we capture their associated risks separately from design risks, even though they are part of the design process.

\paragraph{Performance of model architecture, optimization algorithm, and training procedure} 
Different combinations of model architecture, optimization algorithm, and training procedure have different effects on its final performance (e.g., accuracy, generalization). These choices are independent of \emph{modeling choices} (discussed in \Cref{sec:design}), where the ML practitioner translates a problem statement into an ML problem/task (e.g., by defining the input and output space). For example, a language model can be trained with either the causal or masked language modeling objective \cite{devlin2019bert}. While the latter is suitable for text classification, it may be suboptimal for text generation. Additionally, some training procedures (e.g., domain adversarial training \cite{ganin2016domain}) may improve the ML system's ability to generalize to new domains with minimal extra training data but may hurt performance on the original domain. While accuracy on general benchmark datasets is often used to differentiate models, a better indicator of real-world efficacy is performance on similar applications, due to nuances in the target distribution and the tendency of state-of-the-art models to be optimized for leaderboards \cite{ethayarajh-jurafsky-2020-utility}.

\paragraph{Reliability and computational cost of machine learning component(s) in production}
Beyond efficacy, it is also important to consider the reliability and resource intensiveness of the chosen ML algorithm, model architecture, and optimization technique combination in production scenarios. From an operational standpoint, a highly accurate system that is computationally intensive or failure-prone may be less desirable than a slightly less accurate one without those flaws.

\paragraph{Explainability/transparency}
Algorithmic opacity and unpredictability can pose risks and make it difficult to ensure accountability \cite{Taeihagh2021}. While new mandated levels of transparency and explainability of algorithms are being demanded through the likes of the EU's General Data Protection Regulation (GDPR) to tackle bias and discrimination \cite{goodman2017european}, it can be at times impossible for the experts to interpret how certain outputs are derived from the inputs and design of the algorithm \cite{Taeihagh2021, arrieta2020explainable}. This suggests the difficulty of assigning liability and accountability for harms resulting from the use of the ML system, as inputs and design rules that could yield unsafe or discriminatory outcomes cannot as easily be predicted \cite{kroll2015accountable,kim2017crashed}. Therefore, a system that can explain its decision in the event of a mistake is often desirable in high-stakes applications. A mistake can take the form of an accident resulting from a decision \cite{teslayosemite}, a denied loan \cite{creditdenial2019}, assigning different credit limits based on gender \cite{applecreditcard2019}. While explainability on its own is insufficient to reduce biases in the system or make it safer, it may aid the detection of biases and spurious features, thereby reducing safety and discrimination risks when the flaws are rectified. Other use cases, such as judicial applications \cite{deeks2019judicial}, may \emph{require} such explainability due to their nature. However, not all machine learning algorithms are equal in this regard. Decision trees are often considered highly explainable since they learn human-readable rules to classify the training data,\footnote{It is important to note that while the rules derived by model are transparent to a human, they may not be satisfactory explanations for a wronged stakeholder.} while deep neural networks are a well-known example of a black-box model. While there have been recent advances in explaining neural network predictions \cite{jeyakumar2020can}, researchers have also demonstrated the ability to fool attention-based interpretation techniques \cite{pruthi-etal-2020-learning}. This may allow developers to prevent the network's predictions from being correctly interpreted during an audit. The choice of an ML algorithm and its training method, therefore, affects this aspect of algorithm risk.

\subsection{Training and validation data risk} 

This is the risk posed by the choice of data used for training and validation. Due to their data-driven nature, the behavior of machine learning systems is often heavily influenced by the data used to train them. An ML system trained on data encoding historical or social biases will often exhibit similar biases in its predictions. Separate from the training data, validation datasets are often used to evaluate an ML model's ability to generalize beyond the training data, to new examples from the same distribution, or to examples with different characteristics (other distributions). Representative validation data can be used to detect potential mismatches between the training data and the deployment environment, such as the presence of social biases or spurious features in the training data. We summarize key data risks specific to ML systems and refer the reader to Demchenko \etal\ for a detailed discussion of the general issues around big data \cite{demchenko2013big}.

\paragraph{Control over training and validation data}
Using pretrained models (e.g., GPT-3 \cite{gpt3}, BERT \cite{devlin2019bert}, Inception \cite{Szegedy_2016_CVPR}) for processing unstructured data such as images and text is becoming increasingly common. While this can significantly improve performance, the trade-off is reduced control over the training data for teams that do not pretrain their own models and simply build on top of publicly released models or machine learning API services (e.g., translation). Given the discovery of systemic labeling errors, stereotypes, and even pornographic content in popular datasets such as ImageNet \cite{tsipras2020imagenet,birhane2021large,northcutt2021pervasive}, it is important to consider the downstream ramifications of using models pretrained on these datasets. The studies mentioned above were performed on publicly available datasets; Birhane \etal\ further highlight the existence of pretrained models trained on private datasets that cannot be independently audited by researchers \cite{birhane2021large}.

\paragraph{Demographic representativeness of training and validation data}
Due to the data-driven nature of machine learning, training an ML system on data that insufficiently represent underrepresented demographics may lead to disproportionate underperformance for these demographics during inference, especially if unaccounted for during model design \cite{buolamwini2018gender,suresh2019framework,hooker2021moving}. This is representativeness in the quantitative sense, of the ``number of examples in the training/validation set'', and the performance disparity can result in allocational harms where the minority demographics have reduced access to resources due to the poorer performance. For example, poor automated speech recognition performance for minority dialect speakers (e.g., African American Vernacular English) will have devastating consequences in the courtroom \cite{jones2019testifying,koenecke2020racial}. We may also think of representativeness in the qualitative sense, where stereotypical examples are avoided and fairer conceptions of these demographics are adopted \cite{buolamwini2018gender,garg2018word,khanfaces2021}. Since labels are often crowdsourced, there is the additional risk of bias being introduced via the annotators' sociocultural backgrounds \cite{denton2021whose,arhin2021ground} and desire to please \cite{orne1962social}.

\paragraph{Similarity of training and validation data distribution to deployment distribution} Where demographic representativeness deals with the proportion of subpopulations in the dataset, distributional similarity is more concerned with major shifts between training and deployment distributions. This can occur when there is no available training data matching a niche deployment setting and an approximation has to be used. However, this comes with the risk of domain mismatch and consequently, poorer performance. For example, an autonomous vehicle trained on data compiled in Sweden would not have been exposed to jumping kangaroos. Subsequently deploying the vehicle in Australia will result in increased safety risk from being unable to identify and avoid them, potentially increasing the chance of a crash \cite{volvokangaroo2017}.

\paragraph{Quality of data sources}
The popular saying, ``garbage in, garbage out'', succinctly captures the importance of data quality for ML systems. Common factors affecting the quality of labeled data include annotator expertise level \cite{yan2014learning}, inter-annotator agreement \cite{brants-2000-inter,nowak2010reliable}, overlaps between validation and training/pretraining data \cite{lewis-etal-2021-question}. The recent trend towards training on increasingly large datasets scraped from the web makes manual data annotation infeasible due to the sheer scale. While such datasets satiate increasingly large and data-hungry neural networks, they often contain noisy labels \cite{li2017webvision}, harmful stereotypes \cite{bender2021dangers,abid2021persistent,dodgec42021}, and even pornographic content \cite{aidungeon2021}. Kreutzer \etal\ manually audited several multilingual web-crawled text datasets and found significant issues such as wrongly labeled languages, pornographic content, and non-linguistic content \cite{caswell2021quality}. An even greater concern from the ML perspective is the leakage of benchmark test data and machine-generated data (e.g., machine-translated text, GAN-generated images) into the training set \cite{dodgec42021}. The former was only discovered after training GPT-3 \cite{gpt3}, while the latter is inevitable in uncurated web-crawled data due to its prevalence on the Internet. Researchers have also discovered bots completing data annotation tasks on Amazon Mechanical Turk, a platform used to collect human annotations for benchmark datasets.\footnote{\url{https://twitter.com/RoozbehMottaghi/status/1455563247512211462}} However, cleaning such datasets is no mean feat: blocklist-based methods for content filtering may erase reclaimed slurs, minority dialects, and other non-offensive content, inadvertently harming the minority communities they belong to \cite{dodgec42021}. In fact, the very notion of \emph{cleaning} language datasets may reinforce sociocultural biases and deserves further scrutiny \cite{tan2022thesis}.

\paragraph{Presence of personal information} The presence of personal information in the training data increases the risk of the ML model memorizing this information, as deep neural networks have been shown to do \cite{arpit2017closer,zhang2021understanding}. This could lead to downstream consequences for privacy when membership inference attacks are used to extract such information \cite{shokri2017membership}. We discuss this in greater detail in \Cref{sec:privacy}.

\subsection{Robustness risk}\label{robustness_risk} 

This is the risk of the system failing or being unable to recover upon encountering invalid, noisy, or out-of-distribution (OOD) inputs. There is often significant variation in real-world environments, compared to research benchmarks. For example, objects may appear different under various lighting conditions or wear out over time, and human-generated text often exhibits sociolinguistic variation \cite{labov1973sociolinguistic}. Additionally, malicious actors may exploit flaws in a system's design to hijack it (e.g., in the form of an adversarial attack \cite{goodfellow2015}). The inability to handle the above situations may lead to negative consequences for safety (e.g., autonomous vehicle crashes) or fairness (e.g., linguistic discrimination against minority dialect speakers \cite{tan-etal-2020-morphin,tan-etal-2021-reliability}). Since ML systems sit at the intersection of statistics and software engineering, our definition encompasses two different definitions of \emph{robustness}: the first relates to distributional robustness, where a method is resistant to deviations from the training data distribution \cite{huber2004robust}; the second refers to the ability of a system to ``function correctly in the presence of invalid inputs or stressful environmental conditions'' \cite{8016712}.

\paragraph{Scope of deployment environment}
Similar to \Cref{app_risk}, the deployment environment's scope determines the range of variation the ML system will be exposed to. For example, it may be acceptable for an autonomous robot operating in a human-free environment to be unable to recognize humans, but the same cannot be true for a similar robot operating in a busy town square. A larger range, therefore, usually necessitates either a more comprehensive dataset that can capture the full range of variation or a mechanism that makes the system robust to input variation. A broader scope may also increase the possibility of adversarial attacks, particularly when the system operates in a public environment.

\paragraph{Mechanisms for handling of out-of-distribution inputs}
Out-of-distribution (OOD) inputs refer to inputs that are from a distribution different from the training distribution \cite{hendrycks2017baseline}. They include inputs that \emph{should} be invalid, noisy inputs (e.g., due to background noise, scratched/blurred lenses, typographical mistakes, sensor error), natural variation (e.g., different accents, lens types, environments, grammatical variation), and adversarial inputs (i.e., inputs specially crafted to evade perception or induce system failure). Incorporating mechanisms that improve robustness (e.g., adversarial training \cite{madry2018towards}) reduces robustness risk, but often comes with extra computational overhead during training or inference.

\paragraph{Failure recovery mechanisms}
In addition to functioning correctly in the presence of OOD inputs, system robustness also includes its ability to recover from temporary failure \cite{saxena2017failure}. An example of recovery is an autonomous quadrupedal robot regaining its footing without suffering physical damage after missing a step on the way down a staircase \cite{fankhauser2018robust}.

\subsection{Design risk}\label{sec:design} 

This is the risk of system failure due to system design choices or errors. While the ML model is the core component, we should not neglect the risks resulting from how the problem is modeled as an ML task and the design choices concerning other system components, such as the tokenizer in natural language processing (NLP) systems.

\paragraph{Data preprocessing choices}
ML systems often preprocess the raw input before passing them into their modeling components for inference. Examples include tokenization \cite{mielke2021between}, image transformation, and data imputation and normalization. Additionally, data from multiple sources and modalities (image, text, metadata, etc) may be combined and transformed in ETL (extract, transform, load) pipelines before being ingested by the model. The choices made here will have consequences for the training and operation of the ML model. For example, filtering words based on a predefined list, as was done for Copilot \cite{copilotWord2021}. Such simplistic filtering does not account for the sociolinguistic nuances of slurs and offensive words, and could unintentionally marginalize the very communities it was intended to protect \cite{bender2021dangers}.

\paragraph{Modeling choices}
The act of \emph{operationalizing} an abstract construct as a measurable quantity necessitates making some assumptions about how the construct manifests in the real world \cite{jacobs-wallach2021}. Jacobs and Wallach show how the measurement process introduces errors even when applied to tangible, seemingly straightforward constructs such as height \cite{jacobs-wallach2021}. A mismatch between the abstract construct and measured quantity can lead to poor predictive performance, while confusing the measured quantity for the abstract construct can have unintended, long-term societal consequences \cite{bowker2000sorting}. 

In contrast to recent end-to-end approaches for processing unstructured data (e.g., image, text, audio), ML systems that operate on tabular data often make use of hand-engineered features. The task of feature selection then rests on the developer. Possible risks here include: 1) Training the ML component on spurious features; 2) Using demographic attributes (e.g., race, religion, gender, sexuality) or proxy attributes (e.g., postal code, first or last name, mother tongue) for prediction \cite{oneil2016weapons}. The former could result in poor generalization or robustness, the latter, entrenching discrimination against historically marginalized demographics. For example, the automated essay grading system used in the GRE was shown to favor longer words and essays over content relevance, unintentionally overscoring memorized text \cite{bridgeman2012comparison,ramineni2018understanding}. Other automated grading systems have proven to be open to exploitation by both students and NLP researchers \cite{chingaming2020,ding-etal-2020-dont}.

\paragraph{Specificity of operational scope and requirements}
Designs are often created based on requirements and specifications. Consequently, failing to accurately specify the requirements and operational scope of the system increases the risk of encountering phenomena it was not designed to handle. This risk factor is likely to be most significant for ML systems that are high stakes or cannot be easily updated post-deployment.

\paragraph{Design and development team}
Although software libraries such as PyTorch \cite{paszke2019pytorch} and \texttt{transformers} \cite{wolf2020transformers} are increasing the accessibility of machine learning, a technical understanding of ML techniques and their corresponding strengths and weaknesses is often necessary for choosing the right modeling technique and mitigating its flaws. Similarly, good \emph{system design} requires engineers with relevant experience. A team with the relevant technical expertise may be able to identify gaps in the design requirements and help to improve them. Conversely, the lack of either increases the risk of an ML system failing post-deployment or having some unforeseen effects on the affected community. There have been calls for mandatory certification of engineers to ensure a minimum level of competency and ethical training, though they are largely voluntary \cite{cihon2021ai}. Additionally, the diversity of a team (in terms of demographics) will affect its ability to identify design decisions that may disproportionately impact different demographics \cite{teamdiversity2016}, such as using proxy attributes in modeling or training an international chatbot only on White American English. 

\paragraph{Stakeholder and expert involvement}
Since the development team is unlikely to be able to identify all potential negative consequences, other experts (e.g., human rights experts, ethicists, user researchers) and affected stakeholders should be consulted during the design process \cite{friedman2019value,tan-etal-2021-reliability}. This involvement helps to mitigate the team's blind spots and identify unintended consequences of its design choices, allowing them to be addressed before anyone is harmed. In some cases of participatory machine learning, affected stakeholders can directly influence the system's design as volunteers \cite{halfaker2020ores}.

\subsection{Implementation risk}
This is the risk of system failure due to code implementation choices or errors. A design may be imperfectly realized due to the organization's coding, code review, or code integration practices leading to bugs in the system's implementation. Additionally, the rise of open-source software packages maintained by volunteers (e.g., PyTorch) brings with them a non-trivial chance for bugs to be introduced into the system without the developers' knowledge \cite{thungbugs2012}.

\paragraph{Reliability of external libraries}
Software development is increasingly reliant on open source libraries, and machine learning is no different. Despite their benefits (e.g., lower barrier to entry), using external libraries, particularly when the development team is unfamiliar with the internals, increases the risk of failure due to bugs in the dependency chain \cite{thungbugs2012}. Additionally, over-reliance on open source libraries may result in critical systems going down if the dependencies are taken offline \cite{npm2016}. The level of risk here is therefore determined by the reliability of and community support for the library in question. For example, a library that is widely used and regularly updated by a paid team will likely be more reliable than one released by a single person as a hobby project, even though both are considered open source libraries. However, this is not a given, as the recently discovered Log4j vulnerability demonstrates \cite{log4j2021}. Other common sources of bugs resulting from the use of external libraries are API changes that are not backward-compatible \cite{zhangtfbugs2018}.

\paragraph{Code review and testing practices}
The intertwined nature of the data, model architecture, and training algorithm in ML systems poses new challenges for rigorously testing ML systems \cite{zhang2020machine}. In addition, deep learning systems often fail silently and continue to work despite implementation errors.\footnote{\url{https://ppwwyyxx.com/blog/2017/Unawareness-Of-Deep-Learning-Mistakes}} Good code review and unit testing practices may help to catch implementation errors that may otherwise go unnoticed, lowering the implementation risk \cite{schaul2013unit}.

\subsection{Control risk}\label{control_risk}
This is the difficulty of controlling the ML system. In many scenarios, the ability to shut down an ML system before it causes harm can significantly reduce its second-order risks. For example, the ability to instantly override an autonomous weapon system's decision may be the difference between life and death for a wrongly targeted civilian \cite{firlej2020regulating}.

\paragraph{Level of autonomy}
ML systems are often designed with different levels of autonomy in mind: human-in-the-loop (human execution), human-on-the-loop (human supervision), and full autonomy \cite{nahavandi2017trusted, firlej2020regulating}. Fully autonomous systems may be more difficult to regain control of, in the event of a malfunction; however, it may be simpler to program contingency measures since system developers may assume that the system always bears full responsibility. For example, a real estate company's automated house-flipping system was able to proceed with purchasing over 6,000 houses even after its neural-network--based forecasting algorithm\footnote{\url{https://www.zillow.com/z/zestimate}} generated inaccurate forecasts of house prices during the COVID-19 pandemic, resulting in financial losses of over USD 420,000,000 \cite{zillowbuying2021}. In contrast, its competitor emerged relatively unscathed due to its use of human supervision \cite{redfinzillowbuying2021}.  On the other hand, although a human-supervised system is designed to make intervention easier, the dynamics of human-machine interactions may \emph{increase} the difficulty of determining responsibility as a situation unfolds. While human oversight is theoretically desirable, the above paradox indicates that a human-on-the-loop design could increase control risk if the additional complexity is not accounted for.

\paragraph{Manual overrides}
In human-on-the-loop and fully autonomous systems, the ability to rapidly intervene and either take manual control of or shut down the system is crucial to mitigating the harms that result from misprediction. One factor that significant impacts this ability is the latency of the connection to the ML system (remote vs.\ on-site intervention). This is particularly important in applications that may cause acute physical or psychological injuries, such as autonomous weapons/vehicles and social media bots with a wide reach. Other factors include the ease with which the human supervisor can identify situations requiring intervention and the ease of transitioning from an observer to actor \cite{lucid2021}. These are often tightly connected to the design choices made with regard to the non-ML components of the system. For example, appropriate explainability/interpretability functionality may help the human supervisor identify failures (e.g., when the system's actions and explanations do not align). For high-stakes applications, human supervisors will need to be sufficiently trained (and potentially certified) to react appropriately when they need to assume control.

\subsection{Emergent behavior risk}
This is the risk resulting from novel behavior acquired through continual learning or self-organization after deployment. Although the most commonly discussed ML systems are those trained on static datasets, \footnote{Systems that are continuously \emph{retrained} fall in this category.} there is a paradigm of machine learning known as continuous, active, or online learning. In the latter, the model is updated (instead of retrained) when new data becomes available. While such a paradigm allows an ML system to adapt to new environments post-deployment, it introduces the danger of the ML system acquiring novel undesirable behavior. For example, the Microsoft Tay chatbot, which was designed to learn from interactions with other Twitter users, picked up racist behavior and conspiracy theories within twenty-four hours of being online \cite{msfttay2016}. This paradigm (and associated risks) will likely be most relevant for robots and other embodied agents that are designed to adapt to changing environments \cite{roy2021machine}.

\paragraph{Task type}
The danger of emergent behaviors will likely differ depending on the task the ML system is designed to perform. For example, an NLP system that is mainly in charge of named entity recognition will likely be less dangerous than a chatbot even if both acquire new behaviors through continual learning since the former has a limited output/action space. Novel behavior can also emerge when ML systems interact with each other. This interaction can take place between similar systems (e.g., AVs on the road) or different types of systems (e.g., autonomous cars and aerial drones). This is similar to the idea of swarm behavior \cite{bonabeau1999swarm,schranz2021swarm}, where novel behavior emerges from the interaction of individual systems. While desirable in certain situations, there remains a risk of unintended negative consequences.

\paragraph{Scale of deployment}
The number of deployed systems interacting is particularly relevant to novel behaviors emerging due to self-organization since certain types of swarming behavior may only emerge when a certain critical mass is reached \cite{drozd2016critical}. For example, swarm behavior would be more likely to emerge in vehicular traffic comprising mainly autonomous vehicles surrounding traditional vehicles than vice-versa.

\section{Second-Order Risks}\label{sec:second_order}
Second-order risks result from the consequences of first-order risks and relate to the risks resulting from an ML system interacting with the real world, such as risks to human rights, the organization, and the natural environment. 

\subsection{Safety risk}
This is the risk of direct or indirect physical or psychological injury resulting from interaction with the ML system. By nature, ML systems take away some degree of control from their users when they automate certain tasks. Intuitively, this transfer of control should be accompanied by a transfer of moral responsibility for the user's safety \cite{porter2018moral}. Therefore, a key concern around ML systems has been ensuring the physical and psychological safety of affected communities. In applications such as content moderation, keeping the system updated may involve the large-scale manual labeling and curation of toxic or graphic content by contract workers. Prolonged exposure to such content results in psychological harm, which should be accounted for when assessing the safety risk of these types of ML systems \cite{newton2020youtube,steiger2021}.

First-order risks may lead to safety risk in different ways. For example, poor accuracy may lead to the system failing to recognize a pedestrian and running them over \cite{teslacrashofficers2021}, a melanoma identifier trained on insufficiently diverse data may result in unnecessary chemotherapy \cite{steele2021ai}, or swarming ML systems that endanger human agents (e.g., high-speed maneuvers via inter-vehicular coordination making traffic conditions dangerous for traditional vehicles) \cite{xu2019vehicular}. The inability to assume/regain control in time may also result in increased safety risk, (e.g, overriding an autonomous weapon before it mistakenly shoots a civilian) \cite{firlej2020regulating}.

\subsection{Discrimination risk}
This is the risk of an ML system encoding stereotypes of or performing disproportionately poorly for some demographics/social groups. ML systems gatekeeping access to economic opportunity, privacy, and liberty run the risk of discriminating against minority demographics if they perform disproportionately poorly for them. This is known as ``allocational harm''. Another form of discrimination is the encoding of demographic-specific stereotypes and is a form of ``representational harm'' \cite{crawford2017}. The Gender Shades study highlighted performance disparities between demographics in computer vision \cite{buolamwini2018gender} while Bolukbasi \etal\ discovered gender stereotypes encoded in word embeddings \cite{bolukbasi2016}. Recent reporting has also exposed gender and racially-aligned discrimination in ML systems used for recruiting \cite{recruitingbias2018}, education \cite{feathersflawed2019}, automatic translation \cite{translation-arrest2017}, and immigration \cite{passportphotobias2018}. We focus on how discrimination risk can result from first-order risks and refer the reader to comprehensive surveys for discussions on the biases in ML algorithms \cite{sun-etal-2019-mitigating,shah-etal-2020-predictive,blodgett-etal-2020-language,jacobs-wallach2021,mehrabi2021bias}.

There are various ways in which first-order risks can give rise to discrimination risk. For example, facial recognition systems may be misused by law enforcement, using celebrity photos or composites in place of real photos of the suspect \cite{garvie2019frtflawed}. This leads to discrimination when coupled with performance disparities between majority and minority demographics \cite{buolamwini2018gender}. Such disparities may stem from misrepresentative training data and a lack of mitigating mechanisms \cite{shah-etal-2020-predictive}. Insufficient testing and a non-diverse team may also cause such disparities to pass unnoticed into production \cite{teamdiversity2016,ehsan2021explainable}. Finally, even something as fundamental as an \texttt{argmax} function may result in biased image crops \cite{yee2021twittercrop}.

\subsection{Security risk}
This is the risk of loss or harm from intentional subversion or forced failure. Goodfellow \etal\ discovered the ability to induce mispredictions in neural computer vision models by perturbing the input with small amounts of adversarially generated noise \cite{goodfellow2015}. This is known as an \emph{evasion attack} since it allows the attacker to evade classification by the system. Some attacks emulate natural phenomena such as raindrops, phonological variation, or code-mixing \cite{belinkov2018synthetic,tan-etal-2020-morphin,eger-benz-2020-hero,tan-joty-2021-code,zhai2020s}. ML systems tend to be highly vulnerable if the models have not been explicitly trained to be robust to the attack. 

Another attack vector involves manipulating the training data such that the ML system can be manipulated with specific inputs during inference, (e.g., to bypass a biometric identification system) \cite{chen2017targeted}. This is known as "data poisoning." The application, control over training data, and model's robustness to such attacks are potential risk factors.

Finally, there is the risk of model theft. Researchers have demonstrated the ability to ``steal'' an ML model through ML-as-a-service APIs by making use of the returned metadata (e.g., confidence scores) \cite{tramer2016stealing,orekondy2019knockoff,krishna2019thieves,keskar-etal-2020-thieves}. Extracted models can be deployed independent of the service, or used to craft adversarial examples to fool the original models. The application setting and design choices significantly affect the amount of metadata exposed externally. For example, while an autonomous vehicle does not return the confidence scores of its perception system's predictions, model thieves may still be able to physically access the system and directly extract the model's architecture definition and weights.

\subsection{Privacy risk}\label{sec:privacy}
The risk of loss or harm from leakage of personal information via the ML system. \samsonedit{Although we only focus on privacy in this section, w}e use the GDPR's definition of personal data \samsonedit{due to its broad coverage}: ``any information relating to an identified or identifiable natural person''.\footnote{\url{https://www.gdpreu.org/the-regulation/key-concepts/personal-data}} Privacy breaches often result from compromised databases \cite{equifax2017} and may be mitigated with proper data governance and stewardship \cite{rosenbaum2010data}. However, we wish to highlight privacy risks that are specific to ML systems. Although federated learning \cite{shokri2015privacy} has been proposed to avoid storing training data in a central location (avoiding the problem of compromised databases), it may still be possible to recover training examples from a model learned in this manner \cite{geiping2020inverting,geng2021towards}. Researchers have also demonstrated that information about the training data can be retrieved from an ML model \cite{fredrikson2015model,shokri2017membership,choquette-choo21membership}, and in some cases, the training examples themselves can even be extracted \cite{carlini2020extracting}. Therefore, simply securing the training data is now insufficient.

\subsection{Environmental risk}
The risk of harm to the natural environment posed by the ML system. There are three major ways in which ML systems can harm the environment. The first is increased pollution or contribution to climate change due to the system's consumption of resources. This relates to the energy cost/efficiency during training and inference, hence, the energy efficiency of the chosen algorithm, its implementation, and training procedure are key factors here \cite{lacoste2019quantifying,strubell2019energy,anthony2020carbontracker}. Other key factors include the energy efficiency of the system's computational hardware and the type of power grid powering the ML system since some power sources (e.g., wind turbines) are cleaner than others (e.g. fossil fuels) \cite{henderson2020towards}.

The second is the negative effect of ML system's predictions on the environment and relate to the system's use case, prediction accuracy, and robustness. For example, an ML system used for server scaling may spin up unnecessary resources due to prediction error, causing an increase in electricity consumption and associated environmental effects. Another ML system may be used to automatically adjust fishing quotas and prediction errors could result in overfishing.

Finally, automating a task often results in knock-on effects such as increased usage due to increased accessibility. This is known as the Jevons Paradox \cite{jevons1865coal} or Khazzoom-Brookes postulate \cite{khazzoom1980economic,brooks1990economic,saunders1992khazzoom}. For example, public transit users may adopt private autonomous vehicles and cause a net increase in the number of vehicles on the road \cite{milakis2017policy}.

\subsection{Organizational risk}
The risk of financial and/or reputational damage to the organization building or using the ML system. An organization may incur said damage when said ML system is shown to result in negative consequences for safety, fairness, security, privacy, and the natural environment. For example, a company was lambasted for its search engine's response to a query about India's ugliest language \cite{googlekannada2021}. Reputational damage can also occur if the public \emph{perceives} the system to potentially result in said negative consequences, such as in the case of a police department trialing the Spot robot \cite{nypdspot2018}.

\subsection{Other ethical risks}
Although we have discussed a number of common risks posed by ML systems, we acknowledge that there are many other ethical risks such as the potential for psychological manipulation, dehumanization, and exploitation of humans at scale \cite{harmsmodel2020}. This is aligned with the notion of surveillance capitalism, in which humans are treated as producers of data that are mined for insights into their future behavior \cite{zuboff2015big}. These insights are often used to sell advertisement exposures. This incentive mismatch between the public and companies can lead to design choices that are detrimental to the former but beneficial to the latter \cite{zuboff2019age}. Examples include the fanning of religious tensions that increased offline violence \cite{myanmarfb2018,fbwhistleblower2021} and encouraging the proliferation of outrageous content to increase engagement \cite{angryfb2021}.

\section{Discussion}
The negative impacts of an ML system (e.g., fairness, safety, environmental impact, surveillance capitalism) are often the result of suboptimal design choices during its conception, design, and implementation. Existing work on software RA largely focuses on the risks to the project's success but not the project's risk to the world. Additionally, they were developed for deterministic software systems and do not account for the new risks posed by software systems that learn from data. On the other hand, existing work on algorithmic IA primarily focuses on assessing these impacts or examining the factors leading to specific categories of impacts (e.g., fairness). Existing draft regulations also refer to a vague notion of risk that can be intentionally or unintentionally misinterpreted to suit the actor's position \cite{veale2021demystifying}. To improve the quality of discussions around ML risk, we present a Machine Learning System Risk framework (MLSR) to connect the risk of system failures (first-order effects) to the risk of societal and environmental impacts (second-order effects). We do this by examining advances in the ML literature and connecting them to research on algorithmic impacts.

Drawing connections between specific first- and second-order risks helps ML practitioners pinpoint the parts of the ML system and development process that may have negative impacts. \samsonedit{Although the second-order risks may appear to be significantly more pressing, our framework indicates that they are often the symptoms of problems in the system's design \& development. Hence, addressing the first-order risks in an ML system will naturally mitigate its second-order risks to the external world.} \samsondelete{This allows them to minimize the aforementioned impacts by fixing the offending components.} It is equally critical that both internal and independent regulators holistically assess the risks of proposed and existing ML systems, beyond those often discussed in the ethical AI community. 
Therefore, MLSR is a first step towards a common vocabulary for such assessments. \samsonedit{For practitioners who are looking to reduce the negative impacts of the ML systems they build, we recommend using MLSR to map out how the first-order risks posed by the ML system lead to its second-order risks before using conventional risk estimation methods to quantify them.} Future work may also explore the ways in which it can be combined with RA tools such as severity-likelihood estimation to develop templates for conducting RAs. \samsonedit{Connecting social and environmental impacts to specific technical failures may also help to inspire the creation of new ML techniques that address these risks, creating more diversity in technical research and reducing the emphasis on beating the state-of-the-art.}

\begin{acks}
We are grateful to Anna Bethke, Min-Yen Kan, and Qian Cheng for their insightful feedback on drafts of this paper. Samson Tan is supported by Salesforce and Singapore's Economic Development Board under the Industrial Postgraduate Programme. Araz Taeihagh is supported by the NUS Centre for Trusted Internet and Community through project CTIC-RP-20-02.
\end{acks}

\bibliographystyle{ACM-Reference-Format}
\bibliography{sample-base}

\clearpage
\appendix

\section{Summary of First- and Second-Order Risks}\label{app:tables}
\begin{table}[h]
\footnotesize
    \centering
    \begin{tabular}{l|p{0.55\textwidth}}
         \toprule
         \textbf{First-Order Risks} & \textbf{Risk Factors}\\
         \midrule
         Application &
         \begin{itemize}[leftmargin=*]\vspace{-0.75em}
             \item Application domain
             \item Consequentiality of system actions
             \item Protected populations impacted
             \item Effect on existing power differentials and inequalities
             \item Scope of deployment environment
             \item Scale of deployment
             \item Presence of relevant evaluation techniques/metric
             \item Optionality of interactions
             \item Accountability mechanisms
             \item Stakeholders' machine learning literacy\vspace{-1em}
         \end{itemize}
          \\
          \midrule
         Misapplication & \begin{itemize}[leftmargin=*]\vspace{-0.75em}
             \item Ability to prevent misuse
             \item Ability to detect misuse
             \item Ability to stop misuse\vspace{-1em}
         \end{itemize} \\
         \midrule
         Algorithm & \begin{itemize}[leftmargin=*]\vspace{-0.75em}
             \item Performance of model architecture, optimization algorithm, and training procedure
             \item Reliability and computational cost of machine learning component(s) in production
             \item Explainability/transparency\vspace{-0.75em}
         \end{itemize}\\
         \midrule
         Training \& validation data & \begin{itemize}[leftmargin=*]\vspace{-0.75em}
             \item Control over data
             \item Demographic representativeness of data
             \item Similarity of training,  validation, and deployment distributions
             \item Quality of data sources
             \item Presence of personal information\vspace{-1em}
         \end{itemize} \\
         \midrule
         Robustness & \begin{itemize}[leftmargin=*]\vspace{-0.75em}
             \item Scope of deployment environment
             \item Mechanisms to improve OOD input handling
             \item Failure recovery mechanisms\vspace{-1em}
         \end{itemize} \\
         \midrule
         Design & \begin{itemize}[leftmargin=*]\vspace{-0.75em}
             \item Data preprocessing choices
             \item Modeling choices
             \item Specificity of operational scope and requirements
             \item Design and development team
             \item Stakeholder and expert involvement\vspace{-1em}
         \end{itemize} \\
         \midrule
         Implementation & \begin{itemize}[leftmargin=*]\vspace{-0.75em}
             \item Reliability of external libraries
             \item Code review and testing practices\vspace{-1em}
         \end{itemize} \\
         \midrule
         Control & \begin{itemize}[leftmargin=*]\vspace{-0.75em}
             \item Level of autonomy
             \item Manual overrides\vspace{-0.75em}
         \end{itemize} \\
         \midrule
         Emergent behavior & \begin{itemize}[leftmargin=*]\vspace{-0.75em}
             \item Task type
             \item Scale of deployment\vspace{-0.75em}
         \end{itemize} \\
         \bottomrule
    \end{tabular}
    \caption{Summary of first-order risks and their risk factors}
    \label{tab:first_order}
\end{table}

\begin{table}[t]
\footnotesize
    \centering
    \begin{tabular}{l| l | c}
         \toprule
         \textbf{Second-Order Risks} & \textbf{Common First-Order Risks} & \textbf{Examples}\\
         \midrule
         Safety & Application & Neural generation-based chatbot for dispensing health advice\\
         & Misapplication & Leaving the driver's seat of a semi-autonomous vehicle while in motion\\
         & Algorithm & Poor human identification accuracy in AV perception resulting in accidents \\
         & Training \& validation data & Melanoma identifier trained on non-diverse data leading to unnecessary chemotherapy \\
         & Robustness & AV perception system misclassifying the moon as a traffic light while on an expressway\\
         & Design & Under-specification of weather conditions the AV is expected to operate in \\
         & Implementation & Software bugs in open-source library used in AV perception system \\
         & Control & Inability to prevent an autonomous weapon from shooting a civilian\\
         & Emergent behavior & Microsoft Tay learning insults from Twitter users\\
         \midrule
         Discrimination & Application & Prediction of recidivism risk\\
         & Misapplication & Using celebrity photos for identifying criminals using facial recognition system \\
         & Algorithm & Disproportionately poor performance for minority dialect speakers \\
         & Training \& validation data & Training automated immigration control gantry on only White facial features \\
         & Robustness & Poor performance for second-language speakers in interactive voice response systems  \\
         & Design & Inability to recognize dark-skinned individuals due to non-diverse development team\\
         & Control & Inability to detect and shut down a biased ML system \\
         \midrule
         Security & Application & Autonomous weapon vs.\ autonomous vacuum cleaner\\
         & Training \& validation data & Easy access to training data leading to ease of poisoning it\\
         & Robustness & CV-based access control system that is vulnerable to adversarial examples \\
         & Design & Returned confidence scores increasing chances of model theft \\
         & Implementation & Software vulnerability in external library that opens the ML system to hijacking \\
         \midrule
         Privacy & Application & Authentication system requiring implicit/explicit storage of biometric data \\
         & Algorithm & Using ML algorithms that are vulnerable to data extraction \\
         & Training \& validation data & Training an ML system on personal information \\
         & Design & Returned confidence scores increasing chances of membership inference attacks \\
         & Control & ML model stored on a robot that can be physically stolen and hacked \\
         \midrule
         Environmental & Application & Poor accuracy for seasonal fishing quota estimation\\
         & Algorithm & Significant computational resources required for good performance\\
         & Design & High computational resource requirements \\
         & Implementation & Energy-inefficient implementation running on infrastructure powered by fossil fuels\\
         \midrule
         Organizational & All & Reputation  loss from perception of inappropriate applications / malfunctioning systems \\
         \bottomrule
         \end{tabular}
    \caption{Summary of second-order risks and first-order risks that may lead to each risk.}
    \label{tab:second_order}
\end{table}

\end{document}